\documentclass[prb,onecolumn,amsmath,amsymb]{revtex4}

\usepackage{graphicx}
\usepackage{dcolumn}
\usepackage{bm}

\usepackage[usenames]{color}

\begin{document}

\title{The RKKY Coupling between Impurity Spins in Graphene Nanoflakes}

\author{Karol Sza{\l}owski}
\email{kszalowski@uni.lodz.pl}
\affiliation{%
Department of Solid State Physics, University of \L\'{o}d\'{z},\\
ulica Pomorska 149/153, 90-236 \L\'{o}d\'{z}, Poland
}%

\date{\today}

\begin{abstract}
We calculate the indirect charge carrier mediated Ruderman-Kittel-Kasuya-Yosida (RKKY) interaction between magnetic impurities for two selected graphene nanoflakes containing four hexagonal rings in their structure, differing by their geometry. We describe the electronic structure of either charge-neutral or doped nanoflakes using the tight-binding approximation with the Hubbard term, which is treated within the molecular-field approximation. We find pronounced differences in the RKKY coupling energies, dependent on the placement of the pair of magnetic moments in the nanostructure and on the edge form. For an odd total number of electrons in the structure, we predict in some circumstances the existence of ferromagnetic coupling with leading first-order perturbational contribution, while for an even number of charge carriers the usual, second-order mechanism dominates. Therefore, doping of the nanoflake with a single charge carrier is found to be able to change the coupling from an antiferromagnetic to a ferromagnetic one for some geometries.  
\end{abstract}

\maketitle

\section{Introduction}
Two-dimensional graphene, being one of the most promising contemporary materials \cite{Geim1,Geim2} offers unique physical properties due to its electronic structure.\cite{Novoselov1,CastroNeto1,Abergel1} Since its discovery it has attracted concerted theoretical and experimental efforts aimed at understanding its rich physics and making basis for future electronics. One of the directions is focused on integrating the charge and spin degrees of freedom, to create novel spintronic devices.\cite{Rocha1,Yazev2,Zhang1} This encourages studies of magnetic properties of graphene. 

Recently, increasing interest has been focused on the graphene nanostructures.\cite{Snook1,Raeder1,Fernandez1} Such nanostructures can be engineered to merge the properties of graphene with ultrasmall sizes, and can serve as possible building blocks of novel devices (see for example Refs.~\onlinecite{Ezawa1,Ezawa3,Ezawa4,Krompiewski2}), especially in context of transport properties, showing giant magnetoresistance.\cite{Krompiewski1} Significant efforts are made to understand magnetism emerging in finite-size graphene structures without magnetic impurities introduced (an extensive review of that particular topic is presented in the Ref.~\onlinecite{Yazyev1}). However, the possibility of introducing magnetic impurity atoms to the graphene lattice is also taken into account.\cite{Santos1, Wu1, Kang1}

Among the topics attracting the interests of researchers, the problem of indirect magnetic coupling between magnetic impurity spins in graphene, mediated by charge carriers, is worth mentioning. This kind of interaction, known as Ruderman-Kittel-Kasuya-Yosida (RKKY) coupling,\cite{Ruderman1,Kasuya1,Yosida1} can be expected to show unique properties in graphene, different from the behaviour in two-dimensional metals,\cite{BealMonod1} owing to the peculiar, linear dispersion relation for the charge carriers in the vicinity of Dirac points. Various calculations of RKKY interaction in graphene sheets are present in the literature, exploiting the bipartite nature of the graphene crystalline lattice in various (mainly perturbational) approaches.\cite{Vozmediano1,Dugaev1,Cheianov1,Saremi1,Bunder1,Brey1,Sherafati1,Sherafati2,Kogan1} Moreover, tight-binding calculations in real space were performed for such a system.\cite{Annica1,Annica2} Recently, also the problem of two Kondo impurities attracted some attention. \cite{Uchoa1, Hu1}

However, the main efforts have so far focused on the calculations of RKKY coupling properties in infinite graphene planes. On the other hand, the coupling in ultrasmall nanoflakes (or nanodisks, \cite{Ezawa1} containing just a few hexagonal rings) can be expected to deviate from the predictions for infinite system, as the dominance of the edge in a nanoflake substantially modifies the electronic structure and severely breaks the translational symmetry. Especially, the peculiar features of the electronic state at the zigzag edge of graphene (or even graphite) are known both from theoretical predictions\cite{Fujita1,Nakada1} and experimental observations for various systems (e.g. Ref.~\onlinecite{Klusek1}), especially quantum-dot systems.\cite{Ritter1} The significance of the edge structure has been already found for example in the transport properties of nanoflakes.\cite{Krompiewski1} Moreover, in ultrasmall molecule-like structures, the existence of discrete electronic states significantly separated in energy allows us to expect a range of novel phenomena. Therefore, a sound motivation appears for studies of RKKY coupling between magnetic moments in nanosized graphene structures, which is the aim of the present work. 

To achieve this goal, we use the real-space tight-binding approximation (TBA) supplemented with the Hubbard term for finite, ultrasmall graphene nanoflakes. We perform the exact diagonalization of the single-particle Hamiltonian resulting from the mean field approximation (MFA). Not limiting the calculations to charge-neutral structures, we take into account the possibility of varying the charge concentration electron by electron (charge doping). One of our goals is to search for the configuration for which the coupling changes its character from ferro- to antiferromagnetic as a result of adding or removing a single electron from the system, in close vicinity to the equilibrium electron concentration. In principle, such a possibility might be opened by placing the nanostructure between two tunneling electrodes, providing a single-electron control of mutual orientation of impurity spins (let us mention that the idea of doping-controlled coupling between magnetic moments in bilayer graphene has been very recently raised in the Ref.~\onlinecite{Killi1}). The phenomenon of Coulomb blockade in some graphene nanostructures with two electrodes was studied in Ref.~\onlinecite{Ezawa2}, while another route to modifying the charge concentration is the absorption of gas molecules (as shown in Ref.~\onlinecite{Schedin1}).

\section{Theory}

The analysis of RKKY interaction in graphene-based structures consisting of $N$ carbon atoms bases on the reliable description of the electronic properties. The most commonly used approach to that problem is the TBA \cite{Wallace1,Reich1,CastroNeto1}. The main advantage of the TBA method is the ability to include the full, realistic band structure of graphene with finite bandwidth, not being restricted to the linear part of the dispersion relation close to Dirac points. Therefore, the use of certain cut-off schemes is not necessary to obtain convergent results and the issue of the possible unphysical modifications of the RKKY range function is absent. Moreover, the TBA facilitates the non-perturbational treatment of the problem of magnetic impurities. In our study we will adopt this approach, taking into account the electronic hopping for nearest-neighbours only.

The total Hamiltonian for the ultrasmall graphene structure with two magnetic impurities can be written as
\begin{equation}
\mathcal{H}=\mathcal{H}_{0}+\mathcal{H}_{C}+\mathcal{H}_{imp},
\end{equation}
and contains the contribution from the TBA hopping term $\mathcal{H}_{0}$, Hubbard term $\mathcal{H}_{C}$ as well as impurity potential $\mathcal{H}_{imp}$.

The TBA term has the following form:
\begin{align}
\mathcal{H}_{0}=&-t\,\sum_{\left\langle i,j\right\rangle,\sigma}^{}{\left(c^{\dagger}_{i,\sigma}\,c_{j,\sigma}+c^{\dagger}_{j,\sigma}\,c_{i,\sigma}\right)}.
\end{align}
Here, $c^{\dagger}_{i,\sigma}$ and $c_{i,\sigma}$ denote creation and annihilation operators for $p^{z}$ electrons at sites $i$ and $j$ in the nanoflake, with spin $\sigma=\,\uparrow\,,\,\downarrow$. The summation over nearest-neighbour sites is denoted by $\left\langle i,j\right\rangle$. The parameter $t$ (usually taken as 2.8 eV) is the hopping integral between nearest neighbours. In the further numerical results, all the energies are normalized to $t$.

In order to incorporate the electron-electron correlations induced by the Coulombic interactions, we include the Hubbard term in the Hamiltonian:
\begin{equation}
\label{eq:hubbard1}
\mathcal{H}_{C}=U\sum_{i}^{}{n_{i,\uparrow}\,n_{i,\downarrow}},
\end{equation}
with the electron number operators $n_{i,\sigma}=c^{\dagger}_{i,\sigma}\,c_{i,\sigma}$.

The Hubbard term, capturing the on-site coulombic repulsion only, neglects the long-range part of the interaction. The usefulness of this model is a subject of debate (see the recent review in Ref.~\onlinecite{Kotov1}). However, it is of noticeably wide use in the theory of carbon (nano)structures; see Refs.~\onlinecite{Jung1, Yazyev1, Potasz1,Fujita1, Fernandez1,Fernandez2} and the recent work on carbon nanotubes in Ref.~\onlinecite{Alfonsi1} Along the lines of the discussion in Ref.~\onlinecite{Alfonsi1}, $U$ should be treated as an effective parameter describing the influence of Coulombic interactions. Its value might result from a competition between on-site repulsion and the interaction between electrons located on different sites (especially nearest neighbour). Also, some recent results evidence suppression of the influence of the long-range interactions \cite{Reed1}.

Contrary to the choice of hopping integrals in the TBA term, the situation with the value of on-site Coulomb repulsion parameter $U$ appears less clear. The values accepted vary between $U=2$ eV \cite{Jung1} and even $U\simeq 10$ eV.\cite{Wehling1} In our study, if we include the Hubbard term, we accept a moderate value of $U/t=1$, close to the choice of Yazyev \cite{Yazyev1} and Potasz \emph{et al.} \cite{Potasz1}.   

The interaction of the $z$-component of the on-site impurity spin $S^{z}_{k}$ (located in at lattice site $k$) with an electron spin $s_{i}^{z}=\left(n_{i,\uparrow}-n_{i,\downarrow}\right)/2$ at the same site is described by the Anderson-Kondo Hamiltonian. For two impurities, located at the sites $a$ and $b$, we have:
\begin{equation}
\label{eq:Himp1}
\mathcal{H}_{imp}=\frac{1}{2}J\left(S^{z}_{a} s^{z}_{a}+S^{z}_{b}s^{z}_{b}\right),
\end{equation}
where $J$ is a spin-dependent impurity potential (contact potential).

Let us note that we select the Ising form of the interaction Hamiltonian [\ref{eq:Himp1}] just for simplicity of the calculations. It leads to the final interaction between magnetic impurities described by $\mathcal{H}^{RKKY}=J^{RKKY}\,S^{z}_{a}S^{z}_{b}$. The usage of the Heisenberg exchange Hamiltonian, of the form $H_{imp}=\frac{1}{2}J\left(\mathbf{S}_{a}\mathbf{s}_{a}+\mathbf{S}_{b}\mathbf{s}_{b}\right)$, would just yield $\mathcal{H}^{RKKY}=J^{RKKY} \mathbf{S}_{a}\mathbf{S}_{b}$, without any modification to the indirect exchange integral itself, which is the only subject of our interest in the present work.  

In order to evaluate the RKKY coupling between the impurities at $T=0$, the ground state energy must be found first in the presence of $n=N+\Delta n$ electrons in the structure, for both parallel and antiparallel orientation of the impurity spins. In undoped graphene each carbon atom donates one $p^{z}$ electron, so that $\Delta n=0$ (i.e. half-filling) characterizes the state of charge neutrality. In general, however, the number of electrons present in the system can vary hypotetically between $0$ (empty energetic spectrum) and $2N$ (completely filled spectrum). We limit our further considerations to $\left|\Delta n\right|\leq 6$, in order not to lose the accuracy of electronic spectrum reproduction by means of the TBA method. The RKKY coupling energy between the impurity spins can be related to the difference of total energies by
\begin{equation}
\label{eq:rkky}
E\left(S^{z}_a=\uparrow,S^{z}_b=\downarrow\right)-E\left(S^{z}_a=\uparrow,S^{z}_b=\uparrow\right)=2S^2 J^{RKKY},
\end{equation}
where the positive value of $J^{RKKY}$ corresponds to ferromagnetic coupling (F) and the negative value to antiferromagnetic coupling (AF). In our case (when electronic structure description involves NN hopping only) the coupling values calculated for $\pm\Delta n$ are identical (i.e., electron and hole doping lead to the same results) due to electron-hole transformation symmetry of the Hamiltonian on the finite bipartite lattice.

Let us note that the indirect charge carrier mediated interaction resulting from our calculations can be not necessarily the usual form of RKKY coupling, which is proportional to the square of the contact potential $J$, as resulting from the second order perturbation calculus. \cite{Kogan1} However, we will use in general the term "RKKY interaction" to name the indirect charge carrier mediated coupling between impurity spins, even with different characteristic features, resulting from the simultaneous presence of other mechanisms.
 
In order to deal with the Hubbard term in the Hamiltonian for quite a large system, we adopt the mean field approximation (MFA), which consists in replacement of the form $n_{i,\uparrow}\,n_{i,\downarrow}\simeq n_{i,\uparrow}\,\left\langle n_{i,\downarrow}\right\rangle+n_{i,\downarrow}\,\left\langle n_{i,\uparrow}\right\rangle-\left\langle n_{i,\uparrow}\right\rangle\,\left\langle n_{i,\downarrow}\right\rangle$. This approach has been shown recently to compare successfully with some exact diagonalization method for graphene nanostructures \cite{Feldner1} especially for $U/t<2$ and has been applied to studies of edge magnetic polarization in graphene sheets \cite{Fujita1,Jung1,Yazyev2,Feldner2} or RKKY in infinite graphene.\cite{Annica2} The approximation leads to the effective Hamiltonian defined in single-particle space. The pair of coupled effective single-particle Hamiltonians, $\mathcal{H}^{MFA}_{\uparrow}$ and $\mathcal{H}^{MFA}_{\downarrow}$, for spin-up and spin-down electrons, is obtained (the Hamiltonians depend on the electronic densities $\left\langle n_{i,\sigma}\right\rangle$ and can be treated self-consistently). 

The Hamiltonian matrix for MFA Hamiltonians can be found in the orthonormal basis of single-electron atomic orbitals $c^{\dagger}_{i,\sigma}\left|0\right\rangle$. Then, $N$ single-particle eigenstates indexed by $\mu$ can be found for each electron spin orientation $\sigma$ in the form of linear combination of atomic orbitals  $\left|\psi_{\mu,\sigma}\right\rangle=\left(1/\sqrt{N}\right)\sum_{i=1}^{N}{\gamma^{\mu}_{i,\sigma} c^{\dagger}_{i,\sigma}\left|0\right\rangle}$. Here, the coefficients $\gamma^{\mu}_{i,\sigma}$ for $\mu=1,\dots,n$ set up an eigenvector of the Hamiltonian matrix corresponding to the eigenvalue $\epsilon^{\mu}_{\sigma}$. 

Let us sort the eigenvalues in an ascending order, so that $\epsilon^{1}_{\sigma}$ is the lowest one, etc. Then, the total electronic density at site $i$ in the presence of $n_{\sigma}$ electrons with spin orientation $\sigma$ in the system in the ground state at the temperature $T=0$ can be calculated as $\left\langle n_{i,\sigma}\right\rangle =\sum_{\mu=1}^{ n^{\sigma}}{\left|\gamma^{\mu}_{i,\sigma}\right|^2}$. The corresponding ground-state energy is

\begin{equation}
\label{eq:energy1}
E\left(S^{z}_a,S^{z}_b\right)=\sum_{\sigma=\uparrow,\downarrow}^{}{\sum_{\mu=1}^{n^{\sigma}}{\epsilon^{\mu}_{\sigma}}}.
\end{equation}
The values of $n^{\uparrow}$ and $n^{\downarrow}$ (which add up to the given value of $n$) should be selected so to minimize the total energy. 

After the calculation, the obtained values of electron densities are substituted back into the MFA Hamiltonians and the numerical procedure is repeated iteratively until the satisfactory convergence of the eigenvalues and eigenvectors is achieved. Then the obtained self-consistent numeric value of total energy can be used to calculate RKKY coupling according to the formula given in Eq.~\ref{eq:rkky}.

\section{The numerical results and discussion}

\begin{figure*}
\includegraphics[scale=0.80]{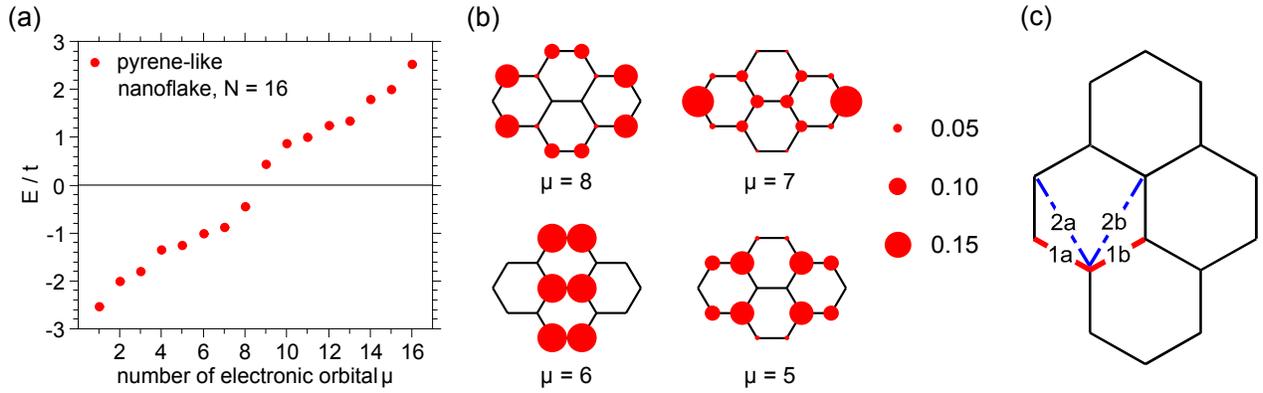}
\caption{\label{fig:N16density} Single-electron (spin-degenerate) energy levels for the pyrene-like graphene nanoflake consisting of 16 carbon atoms, calculated for $U/t=0$ in the absence of magnetic impurities (a). Visualization of electronic densities (probabilities of finding the electron at the given site) corresponding to selected energy levels (b). Schematic view of the nanoflake with two positions of nearest-neighbour impurity ions and two positions of second-neighbour ions, for which the RKKY coupling is discussed (c).}
\end{figure*}

\begin{figure*}
\includegraphics[scale=0.56]{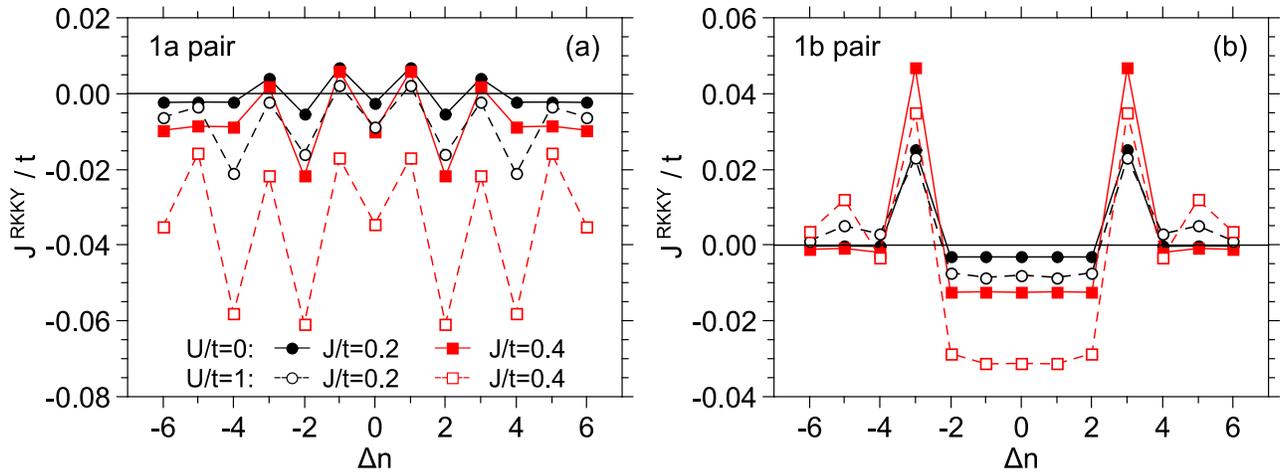}
\caption{\label{fig:N161st} RKKY exchange coupling between selected nearest-neighbour impurity atoms in pyrene-like nanoflake, for impurity positions 1a (a) and 1b (b); see Fig.~\ref{fig:N16density}(c). The values are plotted as a function of the number of electrons in the nanoflake (related to the charge neutrality state). The results are calculated for two values of contact potential, in the presence and in the absence of the Hubbard term.}
\end{figure*}

For our calculations, we selected two graphene nanoflakes, consisting of four hexagonal rings with an even number of carbon atoms. The first one is similar to a pyrene molecule and  contains $N=16$ carbon atoms, and its edge has a zigzag-like character. The second one, triphenylene-like, composed of $N=18$ carbon atoms, possesses the armchair-type edge. Both structures are schematically depicted in Fig.~\ref{fig:N16density}(c) and Fig.~\ref{fig:N18density}(c). 

To study the influence of broken translational symmetry on the RKKY coupling in nanostructures, we calculated $J^{RKKY}$ between two impurities being either nearest or second neighbours, focusing in each case on two different locations of impurity pairs in a given nanostructure. We denote these cases as 1a,1b and 2a,2b, respectively. The considered pairs of magnetic impurities are presented schematically for both graphene nanoflakes in Fig.~\ref{fig:N16density}(c) and Fig.~\ref{fig:N18density}(c). Let us mention that nearest neighbours correspond to two ions in two different sublattices while second neighbours constitute a pair of ions in the same sublattice (referring to the subdivision of the bipartite graphene lattice into A and B sublattices for the finite system). In our calculations, we set $S=1/2$. In the presentation of further results we focus mainly on the influence of charge doping on the characteristics of indirect coupling.

\subsection{Pyrene-like nanoflake}

In order to gain some insight into the electronic structure of the pyrene-like nanoflake, we plot the predicted energy levels resulting from diagonalization of the Hamiltonian in the absence of magnetic impurities and for $U/t=0$ in Fig.~\ref{fig:N16density}(a). The energy states are numbered by $\mu$ and sorted according to ascending energy and each of the states is doubly (spin) degenerate. No additional degeneracy is observed. In the case of charge neutrality the HOMO-LUMO (highest occupied molecular orbital-lowest unoccupied molecular orbital) gap amounts to 0.89 eV (which is in concert with the value resulting from the calculations in Ref.~\onlinecite{Feldner1}, Fig.~2). In order to visualize the electronic densities assigned to the distinct states, we present Fig.~\ref{fig:N16density}(b). There, the values of $\left|\gamma^{\mu}_{i,\sigma}\right|^2$ (probabilities of finding the electron at the given lattice site $i$ for a given state $\mu$) are plotted on the nanoflake scheme, for selected orbitals which are HOMO orbitals for $|\Delta n|\leq 6$. Let us observe that the subsequent states are characterized by significant variability of the corresponding partial charges distribution. If the selected state is occupied only by a single electron, the distribution of partial charge for this orbital reflects also the spin density.

\begin{figure*}
\includegraphics[scale=0.38]{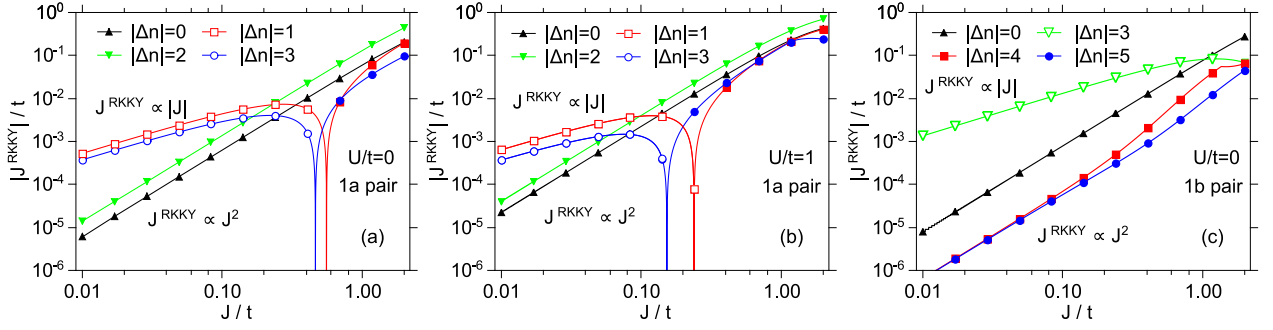}
\caption{\label{fig:N161stlog}Magnitude of RKKY exchange coupling between nearest-neighbour impurities in pyrene-like nanoflake as a function of contact potential, in doubly logarithmic scale. The impurity positions are 1a [(a), for $U/t=0$; (b), for $U/t=1$] and 1b [(c), for $U/t=0$]. The results are presented for selected odd and even electron numbers. Full symbols denote antiferromagnetic coupling, empty symbols - the ferromagnetic one.}
\end{figure*}

\begin{figure*}
\includegraphics[scale=0.43]{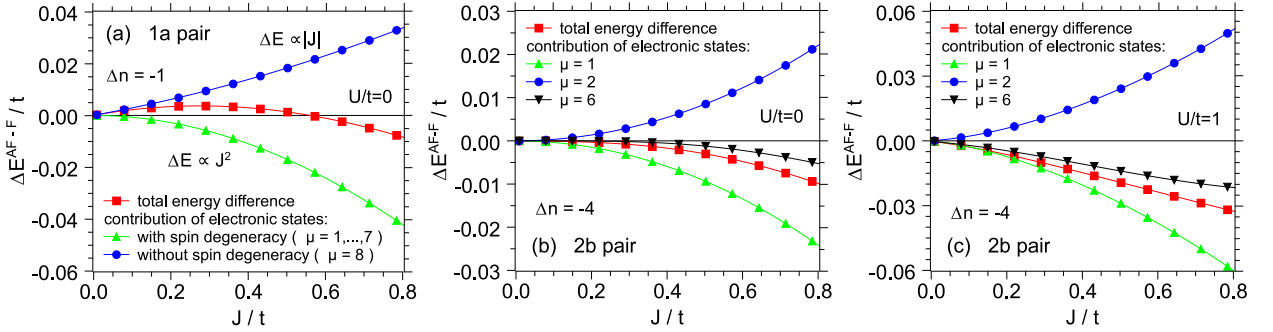}
\caption{\label{fig:N16deltaE2} Energy differences between between antiferro- and ferromagnetically polarized state of magnetic impurities on a pyrene-like nanoflake as a function of the contact potential: 1a pair for $\Delta n=-1$ and $U/t=0$ (a); 2b pair for $\Delta n=-4$ and $U/t=0$ (b) or $U/t=1$ (c). Total energy difference as well as most important contributions are plotted (as described in the plots).}
\end{figure*}

In Fig.~\ref{fig:N161st} we present the values of RKKY exchange integrals calculated for two positions of magnetic impurities, 1a and 1b, as marked in Fig.~\ref{fig:N16density}(c). In order to show qualitatively the influence of increasing contact potential $J$, we prepared the plot for two selected values of $J/t=0.2$ and $0.4$. Moreover, to enable the observation of the Hubbard term effect, we took into consideration the case of $U/t=0$ as well as $U/t=1$. The indirect exchange values are calculated for various deviations in electron number $\Delta n$ from charge neutrality, for $|\Delta n|\leq 6$. As can be observed for both impurity pair positions, the coupling is antiferromagnetic for charge-neutral structure (expected for a bipartite lattice half-filled with the electrons, especially for undoped infinite graphene; e.g. \cite{Annica1}). The interaction is relatively weak and its non-linear rise with increasing $J$ values is also observable. The situation is quite similar for $|\Delta n|=2$, the only difference being the stronger coupling. However, it is quite striking that changing the electronic concentration by one charge carrier from $\Delta n=0$ results in switching of the coupling sign from antiferromagnetic to ferromagnetic, which takes place for $|\Delta n|=3$ as well. In these cases, it can be observed that the change of $J^{RKKY}$ with increasing $J$ is noticeably slower than for even $\Delta n$ values.

\begin{figure*}
\includegraphics[scale=0.56]{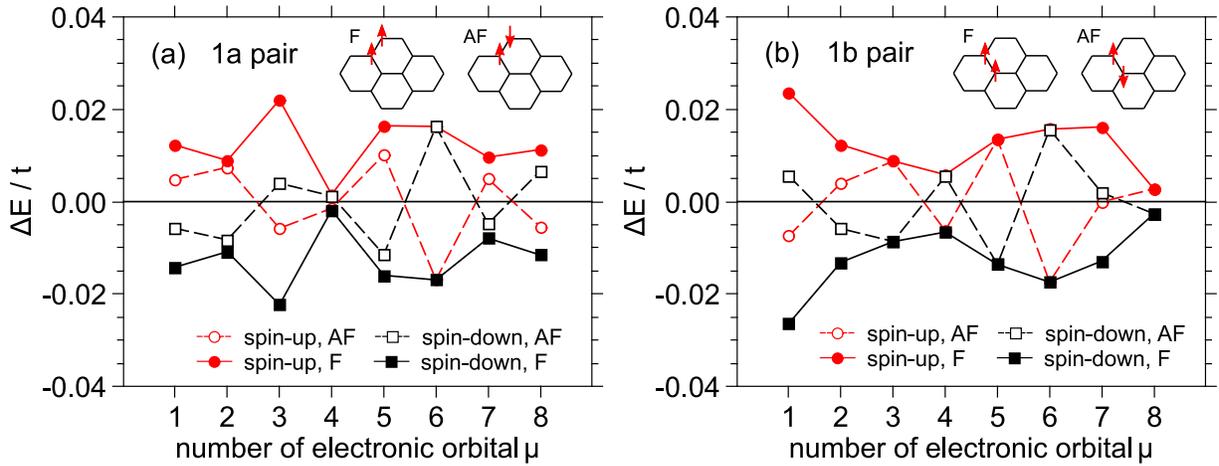}
\caption{\label{fig:N16deltaE} Energy changes for the particular electronic orbitals for spin-up and spin-down electrons in a pyrene-like nanoflake caused by a pair of magnetic impurities, with contact potential $J/t=0.2$. The cases of impurity spins aligned ferro- and antiferromagnetically are presented in each plot (see the insets). The impurity pairs of 1a type (a) and 1b type (b) are discussed.}
\end{figure*}

\begin{figure*}
\includegraphics[scale=0.56]{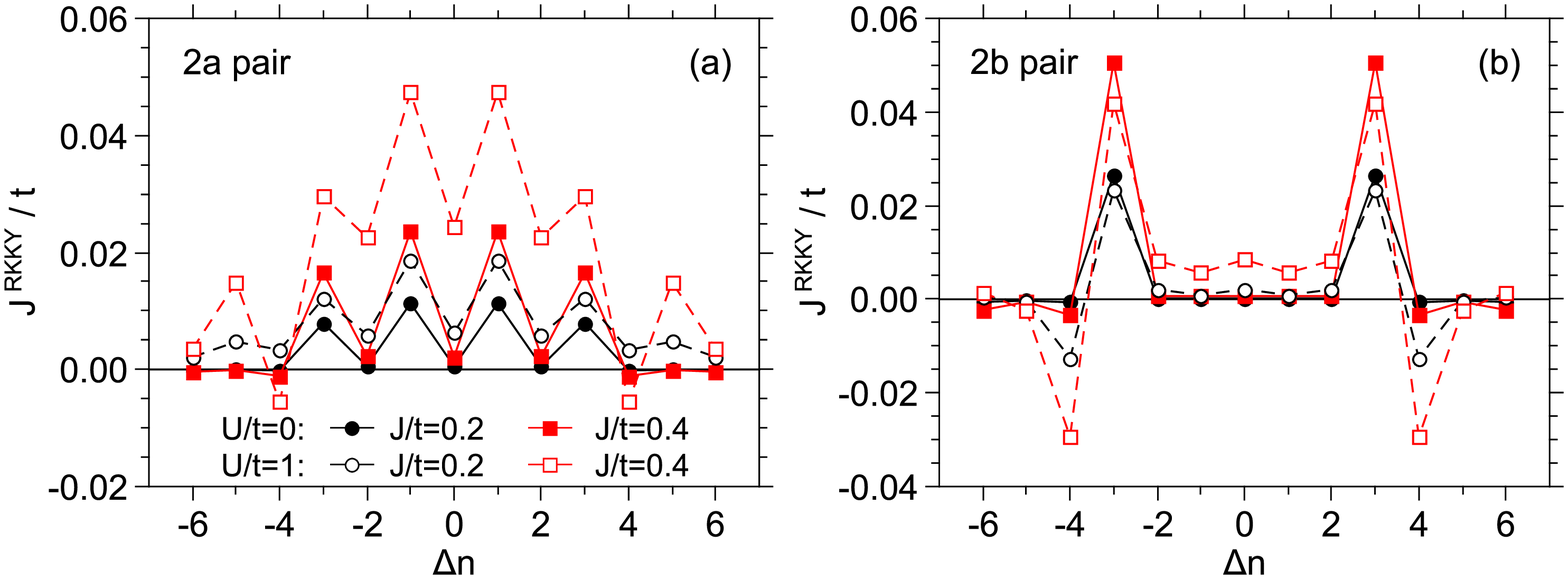}
\caption{\label{fig:N162nd} RKKY exchange coupling between selected second-neighbour impurity atoms in pyrene-like nanoflake, for impurity position 2a (a) and 2b (b); see Fig.~\ref{fig:N16density}(c). The values are plotted as a function of the number of electrons in the nanoflake (related to the charge neutrality).}
\end{figure*}

In order to study more systematically the dependence of exchange coupling on contact potential $J$, we performed the appropriate calculations for nearest-neighbour ions, the results of which are presented in the Fig.~\ref{fig:N161stlog} in double logarithmic scale, which allows us to linearize power-law dependences. The case of 1a impurity pair for $U/t=0$ is illustrated in the Fig.~\ref{fig:N161stlog}(a). It can be verified that two kinds of dependences of $J^{RKKY}$ on $J$ can be distinguished. For selected even total numbers of electrons in the nanoflake ($|\Delta n|=0,2$) the usual dependence $J^{RKKY}\propto J^{2}$ is obeyed, which (according to Fig.~\ref{fig:N161st}) coincides with antiferromagnetic sign of interaction. On the contrary, for odd numbers of electrons ($|\Delta n|=1,3$) we deal with ferromagnetic interaction with $J^{RKKY}\propto |J|$. The latter behaviour tends to convert into to the quadratic dependence on $J$ (and antiferromagnetic exchange) provided that the potential $J$ is strong enough. For clarity, the data for $|\Delta n|=4,5,6$ were omitted, as very close to the results for $\Delta n=0$. Let us also observe that qualitatively the same situation is met in presence of the Hubbard term, for $U/t=1$ [Fig.~\ref{fig:N161stlog}(b)]. It is worth special emphasis that the usual derivation of RKKY exchange integral, within the framework of second-order perturbation calculus, yields always the coupling proportional to the square of the contact potential, which has been recently recapitulated for bipartite graphene lattice. \cite{Kogan1} Therefore, it is of special interest to identify the source of the unusual linear behaviour. Quite similarly, such a behaviour can be observed for the pair of magnetic impurities in location 1b for example when $|\Delta n|=3$ [Fig.~\ref{fig:N161stlog}(c)] (there, the data for $|\Delta n|=0,1,2$ are identical, the same being true for the case of $|\Delta n|=4,6$). Thus, it can be deduced that for selected odd total electron numbers we deal with $J^{RKKY}\propto |J|$.

To investigate more deeply the selected case of two impurities in position 1a and $\Delta n=-1$ and identify the origin of the mentioned unusual behaviour, we plot the energy difference between the AF and F orientations of impurity spins (note that $J^{RKKY}\propto \Delta E^{AF-F}$) as a function of the contact potential $J$ in Fig.~\ref{fig:N16deltaE2}(a). For $\Delta n=-1$, the ground state is characterized by unequal number of spin-up and spin-down electrons. In the plot we resolve the contributions coming from the highest energy orbital occupied by a single electron as well as the total contribution originating from the lower energy orbitals occupied by pairs of electrons of opposite spins. It is noticeable that the latter contribution favours the AF state and is proportional to $J^2$ for not too strong contact potentials (as expected on the basis of second-order perturbation calculus). On the contrary, the orbital occupied by a single electron gives rise to the energy difference which is proportional to $J$ and tends to prefer the F state of the impurities. In the limit of low $J$, the situation is ruled by the singly coupled state so that the total indirect exchange is ferromagnetic and linear in contact potential, while the increase of $J$ leads first to compensation of both contributions and then to domination of the ordinary perturbational $J^2$-proportional behaviour. The linearity of energy difference in $J$ can be explained basing on the fact that for a single electron occupying a given orbital, the leading correction to energy of the orbital [coming from the term Eq.~(\ref{eq:Himp1}) in the Hamiltonian] is of the first- order perturbational kind. If the orbital is occupied by a pair of electrons with opposite spins, first-order corrections for them are of opposite values and cancel each other, while the second-order corrections give rise to the ordinary RKKY interaction. However, when there is no second electron, the uncompensated first-order contribution dominates. Under such circumstances, for F polarization of impurity spins, the first-order correction to the electronic orbital energy due to the presence of two impurities at sites $a$ and $b$ amounts to $\Delta E^{F}=-\frac{1}{2}|J|S \left(|\gamma^{\mu}_{a}|^2+|\gamma^{\mu}_{b}|^2\right)$ (corresponding to the direction of electron spin which minimizes the total energy). Let us assume without loss of generality that $|\gamma^{\mu}_{a}|^2\leq |\gamma^{\mu}_{b}|^2$. If the polarization of impurity spins is AF, then the corresponding correction is $\Delta E^{AF}=-\frac{1}{2}|J|S\left(|\gamma^{\mu}_{b}|^2-|\gamma^{\mu}_{a}|^2\right)$. The resulting energy difference $\Delta E ^{AF-F}=|J|S|\gamma^{\mu}_{a}|^2>0$ clearly makes a ferromagnetic contribution to the indirect coupling. Its magnitude is proportional to the smaller of the electronic densities on the impurity pair sites; thus it is maximized for equal electronic densities on both sites. Such a contribution to indirect coupling yields some resemblance to the double exchange mechanism. \cite{Nolting} However, let us still use the term RKKY interaction to characterize the indirect charge carrier mediated coupling which results from our calculations.

Let us observe, that in some cases we deal with the situation when the electronic density for a given orbital $\mu$ vanishes at least at one of the sites at which the impurity spins are localized. Such an orbital does not indicate any energy difference between AF and F orientation of impurity spins and thus does not give any contribution to the RKKY exchange integral. This is, for example, the case for $\Delta n=-5$ and impurities in the 1a position; see the corresponding electronic densities for the orbital $\mu=6$ in Fig.~\ref{fig:N16density}(b), which is the highest energy orbital occupied by a single electron. Such a situation prevents the indirect interaction from switching to the F sign (like for $|\Delta n|=1,3$), even though the total number of electrons in the system is odd. Quite a similar situation can be observed for impurities in the 1b position, since the electronic density for the orbital $\mu=8$ also vanishes at one of the impurity sites. Therefore, the coupling for $\Delta n=0$ and $\Delta n=-1$ is exactly the same. We note that the value of $J^{RKKY}$ is also unchanged for $\Delta n=-2$, which corresponds to the case when the highest energy orbital $\mu=7$ is occupied by two electrons with opposite spins. However, in that case we can observe that the unperturbed electronic densities are almost equal at both 1b impurity sites. Therefore, the AF state of impurities changes the orbital energy particularly weakly [as can be seen in Fig.~\ref{fig:N16deltaE}(b) for $\mu=7$] and the energy changes for spin-up and spin-down electron for F impurity polarization cancel each other. Therefore, a doubly occupied state $\mu=7$ also gives a particularly weak contribution to exchange integral. This explains the robustness of weak AF RKKY coupling for 1b pair location with respect to deviations from charge neutrality. On the other hand, the same feature of the $\mu=7$ orbital when it is singly occupied gives rise to a particularly strong ferromagnetic contribution to the coupling (seen clearly for $\Delta n=-3$, when $J^{RKKY}\propto |J|$). As mentioned before, this strong ferromagnetic contribution can be explained as a first-order perturbational effect.

To generalize, the possibility of ferromagnetic coupling for nearest-neighbour impurities is open provided that the number of electrons in the system is odd. Under such a condition, the nanoflake gains nonzero magnetic moment, which originates from the HOMO orbital. An additional condition is that the electronic density for the highest energy occupied orbital cannot vanish at both impurity locations. The coupling is particularly enhanced when the electronic densities at the impurity sites are high and close to each other. Then it appears straightforward to explain, that it is energetically favourable to have both impurity spins parallel, as the energy of such a configuration is significantly lowered by the first-order term. Let us notice that in general, for a given weak contact potential $J$, the ferromagnetic couplings, if present, are much stronger than the corresponding antiferromagnetic interactions, which can be seen in Fig.~\ref{fig:N161stlog}.

Let us observe that the presence of a Hubbard term with $U/t=1$ influences remarkably the interaction, especially for the case of charge neutrality, where it leads to strong enhancement of the AF coupling, and eventually it is able to suppress totally any ferromagnetic behaviour for odd electron number, provided that the contact potential is strong enough. This tendency is observable for both nearest-neighbour impurities in the 1a and 1b positions. 

For second-neighbour impurities (results plotted in Fig.~\ref{fig:N162nd}), it is observed that the interaction for both 2a and 2b impurity locations [see Fig.~\ref{fig:N16density}(c)] is almost always ferromagnetic, regardless of the number of electrons, with an exception of $|\Delta n|=4$. Let us mention that the ferromagnetic coupling is expected for an infinite charge-neutral graphene when considering the impurities belonging to the same sublattice, which is the case for second neighbours (e.g. Ref.~\onlinecite{Annica1}). However, for the nanoflakes, pronounced differences emerge depending on the location of the impurity pair. In the case of the 2a location, the coupling is much enhanced for an odd number of electrons (which can be attributed to the same mechanism as that described for F coupling between nearest neighbours; see the linear dependence of $J^{RKKY}$ on $J$ in Fig.~\ref{fig:N162nd} for $|\Delta n|=1,3$). For even values of $\Delta n$ the interaction energy is much weaker. The presence of the Hubbard term with $U/t=1$ tends to build up the coupling. The interaction between impurities in the 2b position exhibits traces of vanishing electronic density of the orbital $\mu=8$ [see Fig.~\ref{fig:N16density}(b)], in analogy to the situation met for the 1b pair. Here, the interaction is strongly damped for $|\Delta n|\leq 2$, while for $|\Delta n|=3$ we observe enhanced F coupling owing to the mechanism mentioned earlier. What is quite interesting, $|\Delta n|=4$ converts the coupling into an antiferromagnetic one.  

In order to comment on the influence of the Hubbard term, we can conclude that its dominant result is to enhance the magnitude of the coupling. The Hubbard term associated energy is lowered when the electronic densities for opposite-spin electrons tend to become unequal, hence acting in hand with the impurity potential in creating an imbalance in spin-up and spin-down electronic densities. This could qualitatively justify why the presence of the Hubbard term increases the magnitude of RKKY coupling, as observed for infinite graphene by Black-Schaffer. \cite{Annica2} Let us exemplify this for the particular case of the 2b impurity pair. The energy differences between the AF and F states of on-site impurities are presented in Figs.~\ref{fig:N16deltaE2}(b) and \ref{fig:N16deltaE2}(c), in the absence and in the presence of Hubbard term, respectively. The total energy difference is plotted together with the separated contribution of the orbitals $\mu=1,2,6$, each occupied by a pair of opposite-spin electrons. The orbitals $\mu=1,2$ are characterized by especially high electronic densities on the impurity sites. Here, all the terms are proportional to $J^2$, i.e., present an ordinary perturbational behaviour for low $J$. As is visible, the energy differences are much more pronounced for all orbitals in the presence of $U/t=1$.

\begin{figure*}
\includegraphics[scale=0.80]{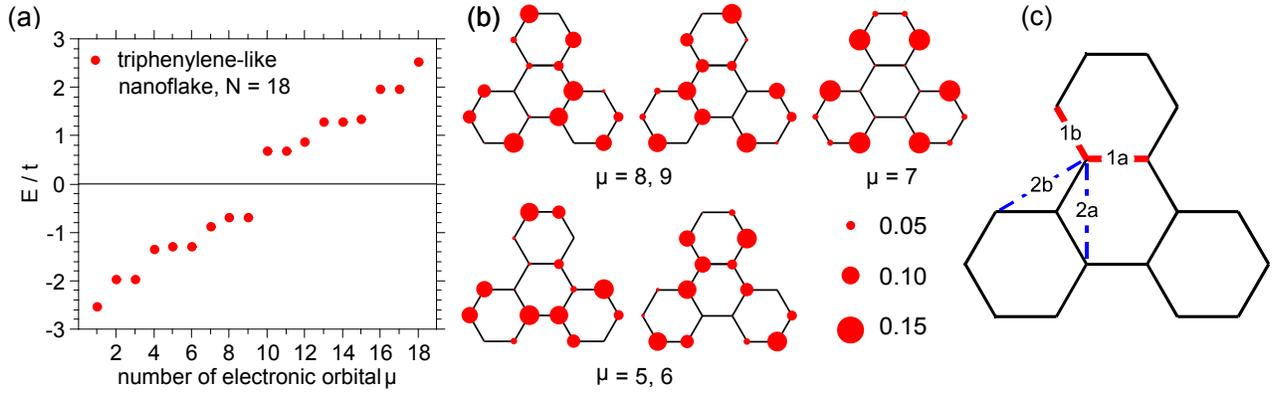}
\caption{\label{fig:N18density} Single-electron (spin-degenerate) energy levels for the triphenylene-like graphene nanoflake consisting of 18 carbon atoms, calculated for $U/t=0$ in the absence of magnetic impurities (a). Visualization of electronic densities (probabilities of finding the electron at the given site) corresponding to selected energy levels (b). Schematic view of the nanoflake with two positions of nearest-neighbour impurity ions and two positions of second-neighbour ions, for which the RKKY coupling is discussed (c).}
\end{figure*}
\begin{figure*}
\includegraphics[scale=0.56]{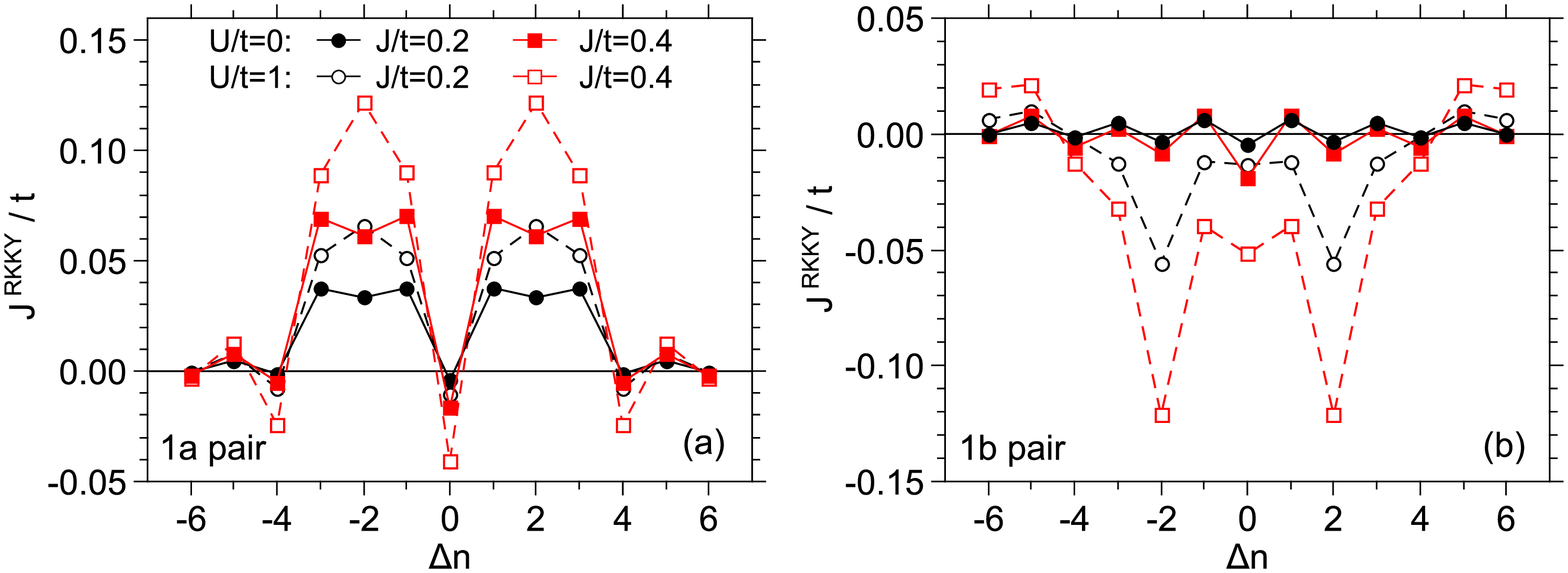}
\caption{\label{fig:N181st} RKKY exchange coupling between selected nearest-neighbour impurity atoms in triphenylene-like nanoflake, for impurity positions 1a (a) and 1b (b); see Fig.~\ref{fig:N18density}(c). The values are plotted as a function of the number of electrons in the nanoflake (related to the charge neutrality).}
\end{figure*}
\begin{figure*}
\includegraphics[scale=0.56]{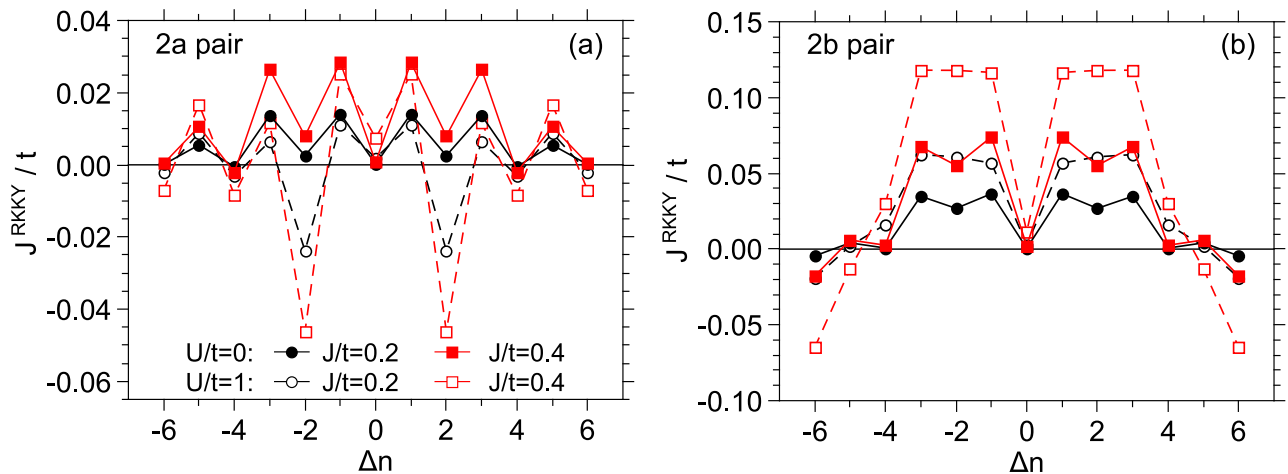}
\caption{\label{fig:N182nd} RKKY exchange coupling between selected second-neighbour impurity atoms in triphenylene-like nanoflake, for impurity positions 2a (a) and 2b (b); see Fig.~\ref{fig:N18density}(c). The values are plotted as a function of number of the electrons in the nanoflake (related to the charge neutrality).}
\end{figure*}

\subsection{Triphenylene-like nanoflake}

The electronic structure of the triphenylene-like nanoflake (obtained from diagonalization of the Hamiltonian in the absence of magnetic impurities and for $U/t=0$) is plotted in Fig.~\ref{fig:N18density}(a). What can be observed is that some states acquire additional two-fold degeneracy, in addition to ordinary spin-degeneracy. For charge-neutral nanoflake we have the HOMO-LUMO gap of 1.37 eV. The electronic densities assigned to the distinct states are presented in Fig.~\ref{fig:N16density}(b) for selected orbitals which constitute HOMO orbitals when $|\Delta n|\leq 6$. There, the values of $\left|\gamma^{\mu}_{i,\sigma}\right|^2$ (partial charges present at the lattice sites for a given state) are plotted on the nanoflake scheme. If the selected state is occupied only by a single electron, the distribution of partial charge for this orbital reflects also the spin density.

Figures~\ref{fig:N181st}(a) and \ref{fig:N181st}(b) show the values of the RKKY coupling between the impurities in two nearest-neighbour positions 1a and 1b [see Fig.~\ref{fig:N18density}(c)]. For an undoped case, an antiferromagnetic interaction is always predicted, again in agreement with expectations for the bipartite lattice. The interesting situation arises for the impurity pair 1b, where the doping with $|\Delta n|=1$ switches the interaction sign to the ferromagnetic one. This sign remains robust against further doping, up to $|\Delta n|\leq 3$, and also the magnitude of coupling varies only weakly. Moreover, the linear dependence of F interaction on $J$ is visible. This phenomenon, present for odd $\Delta n$ values, sparks off from the mechanism described earlier, involving orbitals occupied with a single electron. Here, it is visible in Fig.~\ref{fig:N18density}(b) that the degenerate orbital $\mu=8,9$ is characterized by a large electronic density on both 1a impurity sites. Actually, breaking the symmetry of the nanoflake by introducing the magnetic impurities causes that it is most energetically favourable to have large electronic density at both sites of the 1a pair. What is interesting is that the linear dependence of $J^{RKKY}$ on $J$ survives also for $|\Delta n|=2$. Here, it can be attributed to the fact that for the F polarization of impurity spins, the ground state of the system involves both $\mu=8$ and $\mu=9$ orbitals, each occupied by a single electron, instead of just one doubly occupied orbital. Let us note that the electronic densities for degenerate orbitals $\mu=8,9$ fulfill the condition $|\gamma^{8}_{i,\sigma}|^2+|\gamma^{9}_{i,\sigma}|^2=1/9$, while the orbitals are orthogonal. Therefore, if for one of the orbitals the maximum electronic density at 1a pair sites is achieved, then the second orbital is characterized by vanishing electron density at the same sites. As a consequence, in the F state of the impurities, only one of the orbitals gives contribution to the indirect coupling. The situation is thus analogous to the one expected for the total odd number of electrons. The described occasion of having F coupling for $\Delta n$ even is intimately connected with an additional degeneracy of some electronic orbitals. 

For the 1b pair, the coupling switches its sign from AF to F each time when $\Delta n$ changes from odd to even. The coupling magnitudes are relatively weak. Ferromagnetic couplings are still linearly dependent on $J$. This time, the degeneracy of the orbital $\mu=8,9$ does not play a role at $\Delta n=2$ and the coupling is AF. 

Let us observe that for the magnetic impurities in the 1a position, the presence of the Hubbard term with $U/t=1$ tends to enhance the coupling magnitude, conserving its sign. On the other hand, for the 1b pair, the Hubbard term strongly pushes the interaction toward antiferromagnetic for low $\Delta n$. 

The RKKY exchange integrals calculated for second-neighbour magnetic impurities [in position 2a, 2b; see Fig.~\ref{fig:N18density}(c)] are presented in Fig.~\ref{fig:N182nd}. For 2a placement of the impurities, the interaction is almost always ferromagnetic, but its magnitude is significantly enhanced for odd values of $\Delta n$. For the 2b pair of impurities, the coupling is particularly strong (and ferromagnetic) for $|\Delta n|=1,2,3$, which can be attributed to the same mechanism as described previously in the case of the 1b pair.

\section{Concluding remarks}
We have performed a tight-binding based study of RKKY interaction in two graphene nanoflakes, different in their geometries but both containing four hexagonal rings. We focused on the influence of possible charge doping on the magnitude and sign of an indirect interaction. We also incorporated a Hubbard term in the Hamiltonian to estimate the effect of prototypic Coulomb interaction. In general, we found a pronounced dependence of the RKKY coupling integrals on the location of the pair of on-site magnetic moments in the nanostructure.  

In relation to the odd number of charge carriers in the nanoflakes, we observed a specific contribution to indirect interaction, originating from the highest-energy electronic orbital occupied by a single electron and giving rise to nonzero spin distributed over the structure. This contribution is proportional to the magnitude of the contact potential $J$ and results from an uncompensated first-order correction to the orbital energy due to the presence of a pair of ferromagnetically polarized impurities. This mechanism, which may be regarded as somewhat similar to double exchange, always produces ferromagnetic contribution to indirect exchange and for sufficiently low $J$ it can dominate over the typical contribution proportional to $J^2$ originating from second-order perturbational correction to the energy (being the clue of the ordinary RKKY interaction). Such a ferromagnetic contribution can either change the coupling sign from AF to F for nearest-neighbour impurities or enhance the ferromagnetic coupling between second neighbours, as shown for two types of nanoflakes. The details of the mechanism depend also on the presence or absence of degeneracies in energy spectrum of the nanoflake.

The mechanism leading to the indirect ferromagnetic coupling linearly proportional to $|J|$ might be potentially useful, since it allows switching from AF coupling characteristic of nearest-neighbour impurities for charge neutrality to F coupling by adding or removing a single electron from the nanoflake. Especially advantageous conditions for such a situation occur for a triphenylene-like nanoflake for impurities in the 1a position. What is more, adding one or two further charge carriers does not alter the coupling sign in these particular situations. 

The influence of the Hubbard term is mainly to enhance the magnitude of the coupling, as noted in Ref.~\onlinecite{Annica2} for infinite graphene. Unequal electronic densities for opposite-spin electrons lower the energy coming from the Hubbard term, and the presence of the local spin-dependent impurity potential acts in the same direction. Therefore, the Hubbard term tends to increase the interaction energy in general. On the other hand, the presence of additional long-range repulsion between the electrons residing on different sites should qualitatively tend to compensate the effect of on-site repulsion. This effect could be captured in treating the $U$ parameter not necessarily as a true on-site energy, but rather in the spirit of the effective parameter (as mentioned before, in connection with Ref.~\onlinecite{Alfonsi1})). Sometimes other changes are detected, which can be attributed to the fact that the presence of the Hubbard term itself causes some redistribution of the charge densities over lattice sites with respect to the case when the Coulombic correlations are neglected. These changes also modify the indirect coupling. The application of the MFA approximation to the Hubbard model appears justified for the ground state. \cite{Nolting} However, development of the more accurate approaches (albeit consuming less resources than exact diagonalization) would be valuable with a view to the studies of temperature properties of the nanoflakes. 

The present calculations are performed for $T=0$, exploiting ground-state properties of the system. However, in the ultrasmall, molecule-like structures, the separation of the discrete electronic energy levels is usually of the order of tenths of $t$, where $t=2.8$ eV [see Fig.~\ref{fig:N16density}(c) and \ref{fig:N18density}(c)]. Therefore, it appears that no significant redistribution of the charge carriers between the states can happen due to thermal excitations up to the temperatures of interest (i.e., room temperature). As a consequence, the results for coupling energies appear valid also for nonzero temperatures. Perhaps some finite temperature modifications can be expected in the presence of degenerate spectrum, as the impurity-induced energy splitting between the otherwise degenerate states is considerably small (note the interesting thermodynamics of the undoped metallic-like nanoflakes with degenerate zero-energy states, discussed by Ezawa \onlinecite{Ezawa5}).

In general, we find an immense influence of the electronic structure of the nanostructures on the properties of RKKY interaction, being dependent on the wavefunctions of single orbitals, which are very strongly shaped by the nanoflake geometry. This distinguishes our case from the case of an infinite system, where propagating states are involved, and might open the door for designing the appropriate structures to guarantee the expected indirect coupling features.

\begin{acknowledgments}
The author is deeply grateful to T. Balcerzak for critical reading of the manuscript. 

The numerical calculations have been performed partly with the Wolfram Mathematica 8.0.1 software\cite{Wolfram1} and partly using the \textit{LAPACK} package\cite{laug}.

This work has been supported by Polish Ministry of Science and Higher Education by a special purpose grant to fund the research and development activities and tasks associated with them, serving the development of young scientists and doctoral students.
\end{acknowledgments}


\end{document}